\documentclass[preprint,5p,notitlepage,superscriptaddress,showpacs]{elsarticle}
\usepackage{amsmath,amsfonts,amssymb,verbatim,wasysym}
\usepackage{enumitem}
\usepackage[usenames,dvipsnames]{xcolor}

\journal{Chaos, Solitons \& Fractals}

\begin{document}

\begin{frontmatter}

\title{Aging in coevolving voter models}

\author[inst1,inst2]{Byungjoon Min}
\ead{bmin@cbnu.ac.kr}
\author[inst3]{Maxi San Miguel}
\ead{maxi@ifisc.uib-csic.es}

\affiliation[inst1]{
	organization={Department of Physics and Advanced-Basic-Convergence Research Institute, Chungbuk National University},
  city={Cheongju},
  postcode={28644},
	state={Chungbuk},
  country={Korea}
}
\affiliation[inst2]{
	organization={Department of Medicine, University of Florida},
	city={Gainesville},
	postcode={32610},
	state={FL}, 
  country={USA}
}
\affiliation[inst3]{
	organization={IFISC, Institute for Cross-disciplinary Physics and Complex Systems (CSIC-UIB), Campus Universitat Illes Balears},
  city={Palma},
  postcode={E-07122},
  country={Spain}
}

\begin{abstract}
Aging, understood as the tendency to remain in a given state the longer the persistence time in that state, plays a crucial role in the dynamics of complex systems. In this paper, we explore the influence of aging on coevolution models, that is, models in which the dynamics of the states of the nodes in a complex network is coupled to the  dynamics of the structure of the network. In particular we consider the coevolving voter model, and we introduce two versions of this model that include aging effects: the Link Aging Model (LAM) and the Node Aging Model (NAM). In the LAM, aging is associated with the persistence time of a link in the evolving network, while in the NAM, aging is associated with the persistence time of a node in a given state. We show that aging significantly affects the absorbing phase transition of the coevolution voter model, shifting the transition point in opposite directions for the LAM and NAM.  We also show that the generic absorbing phase transition can disappear due to aging effects.
\end{abstract}


\end{frontmatter}

\section{Introduction}

The interactions among elements of complex systems~\cite{complex,complex_book} can be represented as networks, where the nodes of the network correspond to the individual components of the system and an edge represents an interactions between two nodes \cite{complex_book,barabasi,newman}. In this regard, the dynamics on networks allows us to understand how the structure of these interactions shapes the time evolution of the states of the nodes \cite{newman,boccaletti}. In addition to the dynamics of the state of each node on networked systems, the structure of the underlying network itself can change in time \cite{barabasi,temporal,temporal_book}.
Moreover, many complex networks evolve in response to the influence of their
nodes, often resulting in a coevolution of the nodes' state and the network
structure \cite{zimmermann2004,holme2006,gross2006,vazquez2008,nardini2008,perc2013}.
This interplay between the evolving network structure and the dynamic
behavior of the nodes provides another layer of complexity in many studies of complex systems
such as opinion formation \cite{holme2006,vazquez2008,kimura2008,sudo2013,saeedian,song2023},
epidemic dynamics \cite{gross2006,marceau2010,choi2023}, spin dynamics \cite{mandra2009,raducha_spin},
complex contagion \cite{lambiotte2010,bmin2023}, cultural dynamics \cite{coevAxelrod,centola}, ecological systems \cite{horstmeyer_scirep}, and game-theoretical models \cite{zimmermann2004,ebel2002,eguiluz2005,gonzalez2023}.

Among the various models used to explore coevolutionary behaviors, a coevolving
voter model (CVM) has received much attention due to its simplicity and ability
to capture a variety of phenomena \cite{holme2006,vazquez2008,durrett2012}.
The voter model (VM) \cite{liggett} has
been extensively studied on lattices and static networks with a focus on nonequilibrium ordering dynamics \cite{castellano2003,sood-redner,suchecki,suchecki2005,vazquez2008b,redner2019,kim2024}.
When the network structure itself evolves in response to the states of the nodes, more rich
and complex behaviors can appear, which are not observed in static networks \cite{holme2006,vazquez2008}.
The classical CVM involves two dynamic processes: copying and rewiring.
Each node, in one of two states, can either copy the state
of a randomly selected neighbor or rewire its connection if the
neighbor is in a different state. The relative rate at which these two processes occur is determined by the network plasticity $p$ \cite{holme2006,vazquez2008}.
An interesting feature of the CVM is a generic absorbing phase transition
between a dynamically active phase and a frozen phase which occurs for a critical value of the plasticity.
In a finite system this is manifested as a fragmentation transition between a phase with a single component network with all nodes in the same state, named consensus phase in the context of the voter model,
and a phase with a fragmented network \cite{vazquez2008}.
There have been many variants of the CVM that incorporate multiple different states \cite{shi2013},
noise \cite{diakonova2015}, multilayer coevolution \cite{diakonova2014}, directionality \cite{zschaler},
non-linear interactions \cite{bmin2018,raducha2018,bmin2019,bmin2024}, and
higher-order interactions \cite{horstmeyer2020,papanikolaou}.

Although the CVM effectively captures fundamental coevolutionary dynamics, it overlooks a key time-dependent aspect in dynamics: ``aging.'' Here, aging is understood as the
tendency to remain in a given state the longer the persistence time in that state \cite{tessone,gracia,peralta2020,llabres,baron}. It implies a modification of the dynamics based on the statistics of the persistence times. Aging introduces memory effects in the dynamics
and it is known to play a crucial role in the system evolution \cite{tessone,gracia,artime,abella2022,abella2023}.
In a coevolutionary model, there is a persistence time of the link and a persistence time of the state of a node.
Incorporating aging, the probability of a node changing state or network rewiring decreases
as nodes or links age, that is, as their persistence time increases. This leads to non-Markovian dynamics that deviate from purely memory-less
processes \cite{tessone,gracia,takaguchi,bmin2011,karsai}.
In this work, we consider the role of aging in a CVM considering both the persistence time of a link, that is, the time during which a particular link is present in the network, and the persistence time of the state of the node, that is, the time during which the state of a node has not changed.

The remainder of the paper is organized as follows. Section II introduces the coevolving voter
model with aging effects, detailing the model setup, key assumptions, and the mathematical
formulation of aging in the system. In Sec. III, we examine the analytical predictions
of the CVM with aging focusing on the interevent time distributions. In Sec. IV, we show
results of the CVM with aging, highlighting the impact of aging on the transition between
active and frozen phases. Finally, we summarize our findings and
discuss their implications in Sec. V.

\section{Coevolving voter model with aging}

\begin{figure}
\includegraphics[width=\linewidth]{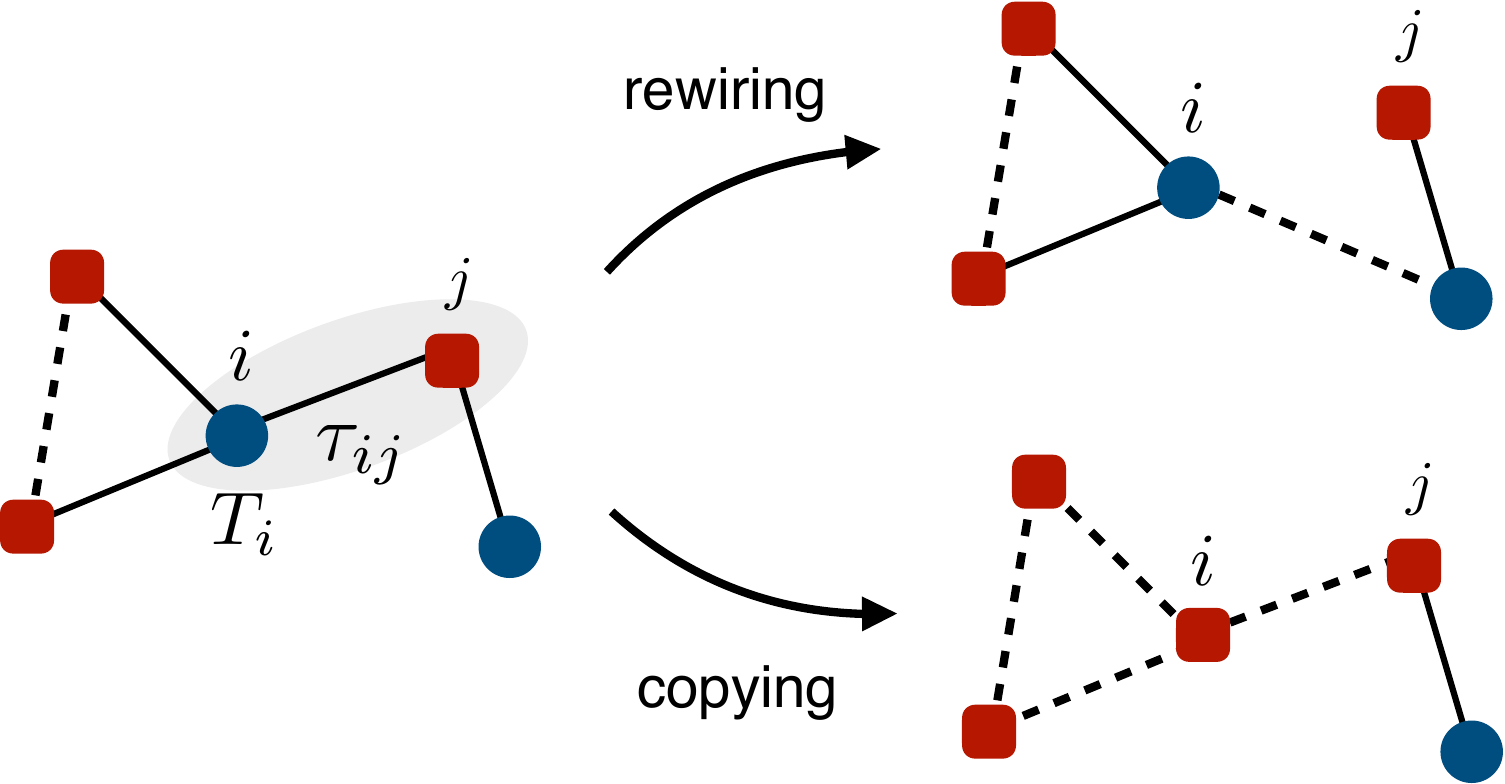}
\caption{
A schematic of a general coevolving voter model with aging is shown.
The model consists of two key processes: copying and
rewiring, with probabilities determined by the node age, $T_i$,
and the link age, $\tau_{ij}$.
}
\label{fig1}
\end{figure}

The CVM is a model used to study the adaptive coupled dynamics of the state of the nodes and the
underlying network structures \cite{holme2006,vazquez2008}.
The model describes a network with $N$ nodes, labeled $i=1,2,\cdots,N$.
We consider binary states of each node, denoted by $s_i(t) \in \{ -1,1\}$
where $t$ is time. The rules of the model are
simple: a node can either adopt the state of one of its neighboring nodes with
probability $1-p$ or cut a connection with a node in a different state and rewire its connection to another node
with probability $p$, where $p$ is the network plasticity.  This intertwined dynamics between state changes and link rewiring generates a
coevolutionary dynamics.

Depending on the value of the plasticity parameter $p$, the system exhibits two
distinct phases in the steady state: active and frozen phases.
The dynamically active phase features a non-zero density of
active links in the thermodynamic limit, which represent links connecting
nodes in different states. Conversely, the frozen phase shows a zero value of
the density of active links, leading to an absorbing phase.
At a critical value $p_c$, a generic absorbing phase transition occurs from
the active to the frozen phase \cite{vazquez2008}. Note that in a finite
system, the active phase, due to
fluctuation effects in finite sizes, manifests itself as a consensus phase in a single component network, while the frozen phase appears as a fragmented network with two disconnected components. Each component has all nodes in the same state, but the state is different for the two disconnected components.

Here we introduce a coevolving voter model with aging defining an age for
each node and link in the network.
The age $\tau_{ij}$ of each link represents the duration for which the link between
nodes $i$ and $j$ has been maintained. In addition, the age $T_i$ of each node
represents the duration for which the current state of node $i$ has been maintained.
The effect of aging is introduced as a time-dependent activation probability that decreases
with the age of the node or the age of the link. Aging can modify the two coupled dynamical processes of the CVM: the dynamics
of the nodes (i.e., state changes) and the dynamics of the links (i.e., network rewiring).
Therefore, we consider two different models depending on whether the CVM dynamics
is modified by link aging or node aging.
In the following subsections, we specify the CVM with i) link aging and ii) node
aging, providing the probabilities of rewiring and copying (see Fig.1).

\subsection{Link Aging Model (LAM)}

We consider here the situation in which
the rewiring probability of the CVM 
decreases as the age of a link increases. To be specific, a link aging model (LAM) is defined by the following rules:
\begin{enumerate}[noitemsep]
\item Initially, the states of nodes are distributed uniformly at random. The age of each
node and each link is set to be unity at $t=0$.
\item At each step, node $i$ and one of its neighbor $j$ are selected at random.
\item Perform one of the three following actions.
\begin{enumerate}
\item With probability $1-p$, node $i$ copies the state $s_j$ of node $j$.
\item With probability $p/{\tau_{ij}^{\alpha_l}}$, if nodes $i$ and $j$ are in different states, node $i$ cuts the link to $j$ and establishes a new link with a random node in the  network
that shares the same state $s_i$ of node $i$. Note that the age of the newly constructed link is set
to one. The parameter $\alpha_l$ controls the influence of aging in the link. When $\alpha_l=0$ no link aging effect is taken into account.
\item With the complementary probability $p(1-1/\tau_{ij}^{\alpha_l})$, nothing happens.
\end{enumerate}
\item After performing this process $N$ times, one unit of time passes, and the age
of each link increases by one.
\item Repeat step (2-4).
\end{enumerate}

\subsection{Node Aging Model (NAM)}

We consider here the situation in which the probability for a node to change state in the CVM 
decreases as the age of the node increases.
To be specific,  a node aging model (NAM) is defined by the following rules:
\begin{enumerate}[noitemsep]
\item Initially, the state of nodes are distributed uniformly at random. The age of each
node and each link is set to be unity at $t=0$.
\item At each step, node $i$ and one of its neighbor $j$ are selected at random.
\item Perform one of the three following actions.
\begin{enumerate}
\item With probability $(1-p)/T_i^{\alpha_n}$, node $i$ copies the state $s_j$ of node $j$.
Note that the age of the node with the changed state is set to one.
The parameter $\alpha_n$ controls the influence of aging in the node. When $\alpha_n=0$ no node aging effect is taken into account.
\item With probability $p$, if nodes $i$ and $j$ are in different states, node $i$ cuts the link to $j$ and establishes a new link with a random node in the  network
that shares the same state $s_i$ of node $i$.
\item With the complementary probability $(1-p)(1-1/T_i^{\alpha_n})$, nothing happens.
\end{enumerate}
\item After performing this process $N$ times, one unit of time passes, and the age
of each node increases by one.
\item Repeat step (2-4).
\end{enumerate}

\setlength{\tabcolsep}{2pt}
\begin{table}
\begin{center}
\begin{tabular}{ |c|c|c|c| }
\hline
\rule{0pt}{3ex}  & copying  & rewiring & nothing happening \\
\hline
\rule{0pt}{3ex} LAM  & $1-p$             &  $p /\tau_{ij}^{\alpha_l}$  & $p(1-1/\tau_{ij}^{\alpha_l})$ \\
\hline
\rule{0pt}{3ex} NAM  & $(1-p)/ {T_i}^{\alpha_n}$ &         $p$     &   $(1-p)(1-1/T_i^{\alpha_n})$   \\
\hline
\end{tabular}
\caption{Model descriptions with the values of parameters}
\end{center}
\end{table}

The summary of the LAM and NAM for the copying and rewiring processes is shown in Table 1.
In the LAM, the probability of copying a link is given by $1-p$, while
the probability of rewiring is proportional to $p/\tau_{ij}^{\alpha_l}$, where
$\tau_{ij}$ is the age of the link between nodes
$i$ and $j$, and $\alpha_l$ controls the influence of aging on the link.
On the other hand, the NAM focuses on node aging, where the probability of copying is
$(1-p)/T_i^{\alpha_n}$, where $T_i$ is the age of node $i$, and
$\alpha_n$ determines the effect of aging. Both models highlight how aging
at the link or node level influences network dynamics through
the processes of copying and rewiring. The role of $\alpha_l$ and $\alpha_n$
is transparent. When $\alpha_n=\alpha_l=0$, there is no effect of aging. In general
the larger the value of $\alpha_n$ or $\alpha_l$, the greater the effect of aging in the dynamics.

\section{Interevent time distributions of the CVM with aging}

In our models, the rate of events in which a node changes state or a link is rewired, can decrease due to aging, leading
to non-Poissonian processes, that are characteristic of empirical human activity patterns \cite{karsai,artime}.
The effects of aging on these models can be understood through a theoretical framework based on the
interevent time (IET) distributions \cite{gracia}.
Let us start with a simple non-Poissonian process involving an age-dependent rate of occurring events.
We will discuss its applications to the LAM and NAM models in the subsequent subsections.

We first define the probability $q(t)$ that an event, which is influenced by aging, occurs at time $t$. To be specific, in the LAM, $q(t)$ corresponds
to the probability that a selected link will be rewired, while in the NAM, $q(t)$  represents
the probability that a selected node will copy a neighbor’s state. As aging progresses,
$q(t)$ decreases over time, leading to a suppression of these processes.
The form of $q(t)$ determines the distribution $M(t)$ of interevent
times (IET), which represents the probability that an event occurs $t$ time steps after
the occurrence of the last event.
For a constant probability of activation implying a constant value of $q(t)$
for all values of $t$, the IET distributions are expected to be exponential distributions
because they follow a Poissonian process \cite{vankampen}. This is the case in our model
when  $\alpha_n=0$ and $\alpha_l=0$. However, when we introduce aging ($\alpha > 0$), the IET
distribution $M(t)$ deviates from exponential distributions.

Next, we find a relation between the probability $q(t)$ and the IET distribution $M(t)$.
The probability that an event does not occurs for $t-1$ time steps is $1 - \sum_{j=1}^{t-1} M(j)$,
and the probability that an event occurs at exactly $t$ time steps is $q(t)$.
This leads to the relationship
\begin{align}
\left( 1 - \sum_{j=1}^{t-1} M(j) \right) q(t) = M(t),
\end{align}
with the initial condition $q(1) = M(1)$. By taking the continuous limit
and expressing this equation in terms of the cumulative IET distribution,
$C(t) = 1 - \int_{1}^{t} M(t') dt'$, we can derive the differential equation
\begin{align}
\frac{1}{C(t)}\frac{d C(t)}{dt} = -q(t).
\label{deq}
\end{align}
This equation shows that the choice of the activation probability $q(t)$
directly determines the form of the cumulative IET distribution, and hence
the IET distributions.

Consider a case where the activation probability is given by $q(t) = b/t^\alpha$
with $b>0$, i.e, $b=1$ in the LAM ($\alpha=\alpha_l$) and NAM model ($\alpha=\alpha_n$).
Then, the solution for Eq.~\ref{deq} can be obtained for the given $q(t)$ and
initial condition $C(1)=1$ as
\begin{align}
C(t) =
\begin{cases}
\exp\left[ \frac{b(t^{1-\alpha}-1)}{\alpha-1}\right], & \text{if } \alpha \neq 1, \\
t^{-b}, & \text{if } \alpha = 1.
\end{cases}
\end{align}
The IET distribution can be straightforwardly obtained from the cumulative
distribution by the relation $M(t)=-dC(t)/dt$:
\begin{align}
M(t) =
\begin{cases}
b \ t^{-\alpha} \exp\left(\frac{b( t^{1-\alpha}-1)}{\alpha-1}\right), & \text{if } \alpha \neq 1, \\
b \ t^{-1-b}, & \text{if } \alpha = 1.
\end{cases}
\end{align}
For $\alpha=0$, the IET distribution decays as an exponentially function $b e^{-b(t-1)}$, and
for $\alpha=1$, it simplifies to a power-law decay with exponent $(-1-b)$.

\begin{figure}
\includegraphics[width=\linewidth]{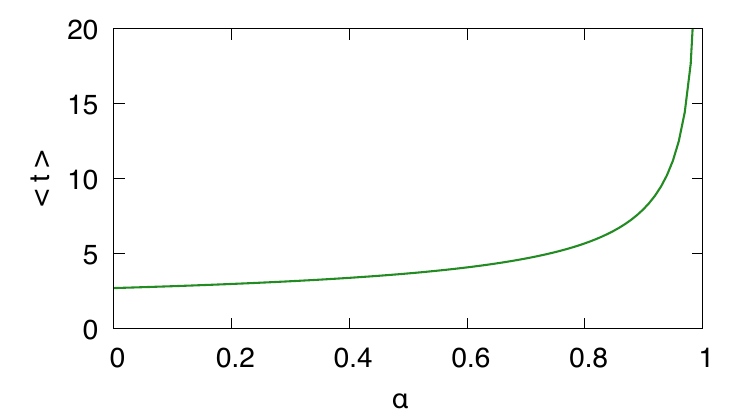}
\caption{
Average interevent time $\langle t \rangle$ as a function of $\alpha$ is depicted, based on Eq.~(5).
}
\label{fig_time}
\end{figure}

From the IET distribution, we can obtain an approximation for the average interevent time in the continuum limit
as $\langle t \rangle \approx \int_{t=0}^\infty t M(t) dt$ .
In our model, we set $b=1$ and the average interevent time is
\begin{align}
\langle t \rangle =
\begin{cases}
[(1-\alpha)e]^{\frac{1}{1-\alpha}} \ \Gamma \left(\frac{\alpha-2}{\alpha-1} \right), &  \text{if } \alpha < 1, \\
\infty, & \text{if } \alpha \ge 1,
\end{cases}
\label{eq:mean}
\end{align}
where $\Gamma(x)$ is the gamma function.
Depending on the value of $\alpha$ which controls the aging function, we expect two different outcomes as shown in Fig.~\ref{fig_time}. When $\alpha < 1$, the average interevent time $\langle t \rangle$
converges to a finite value, indicating that the system evolves with
a characteristic timescale. In this case, events occur with a longer interevent time than
the dynamics without aging $\alpha=0$, leading to slow down.
On the other hand, when $\alpha \ge 1$, infinitely long interevent times are expected.
For instance, for the ordinary voter model with aging it is known \cite{peralta2020} that the
system for $\alpha<1$ approaches the consensus phase with exponential ordering dynamics.
For $\alpha>1$ the system does not order, but for the marginal case $\alpha=1$ \cite{gracia}
it approaches the consensus phase with the density of
active links decaying as a power-law in time. Therefore in the following we restrict most of our analysis to the cases $\alpha \leq 1$.

\subsection{Interevent time distributions of the LAM}

To investigate the effect of aging in the CVM, we must account for the coevolutionary
nature of the system, where the probabilities of copying and rewiring are intertwined.
For general cases of
rewiring probability $0<p<1$, the coevolutionary dynamics makes exact analytical approaches intractable . However, in the special cases where $p=0$ (only copying occurs) or
$p=1$ (only rewiring occurs), the two processes are decoupled, allowing for an analytical approach.
Additionally, while it is generally intractable to derive the inter-event time distributions for  $0<p<1$, we
can still estimate the effect of the aging based on the results for the two limiting cases.

Let us consider the LAM. In this model, the process of flipping the state of nodes occurs
at a constant rate of $1-p$, like in the standard CVM. Aging  plays a role only for
the link rewiring process, which happens with probability $p/\tau_{ij}^{\alpha_l}$.
We begin by examining the two extreme
cases where $p=0$ and $p=1$. In the case of $p=0$, the LAM is simply reduced to
the original VM \cite{sood-redner,vazquez2008b}.
Conversely, when $p=1$, only link rewiring occurs without node state dynamics. Therefore, once all
the initial active links are removed through rewiring, the system reaches a frozen state.
Due to the aging effect, the interevent time for occurring link rewiring varies according
to the aging exponent $\alpha_l$. The average interevent time
is simply given by Eq.~\ref{eq:mean},
which shows that as $\alpha_l$ increases, the link rewiring process slows down. In particular,
when $\alpha_l \ge 1$, the mean interevent time diverges and the system does not reach
a frozen state within a finite time in the thermodynamic limit $N \rightarrow \infty$.
When $0 < p < 1$, link rewiring is coupled with the node-copying process,
so that the exact analytical calculation is intractable.
However, in this case, link aging still slows down the link rewiring, so that copying processes become relatively more frequent.

\begin{figure*}
\includegraphics[width=\linewidth]{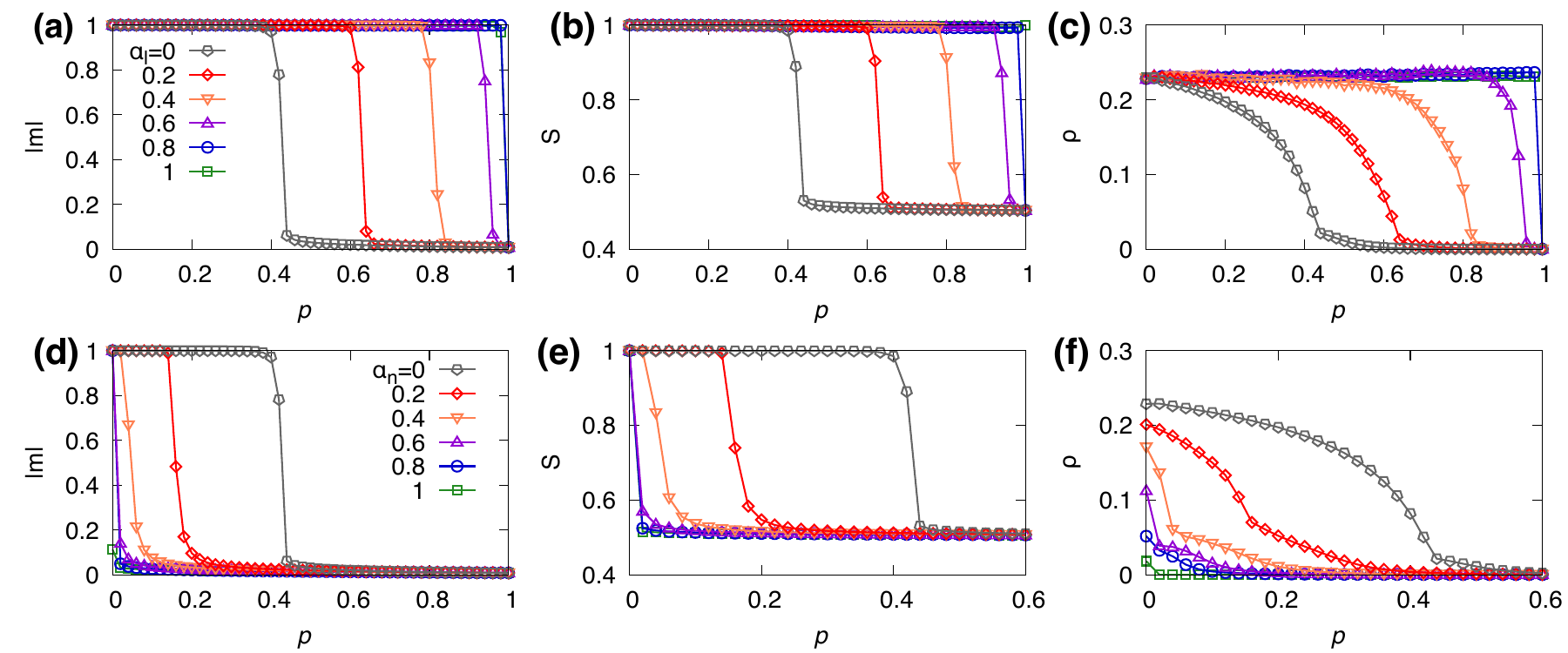}
\caption{
LAM:(a) The absolute value of magnetization $|m|$, (b) the size of the largest component $S$,
(c) the fraction of active links in survival runs $\rho$ as a function of $p$ for various $\alpha_l$ for
the link aging model, and  NAM: (d) $|m|$, (e) $S$, (f) $\rho$ for various $\alpha_n$ for
the node aging model are shown.
We use random regular networks with the average degree $z=4$ and $N=10^4$.
}
\label{fig3}
\end{figure*}

\subsection{Interevent time distributions of the NAM}

In the NAM, aging affects the probability that a node changes its state. This probability decreases with the persistence time of the node in the same state. Aging slows down the occurrence of copying processes as predicted by the longer average interevent times.
The slowing down of the copying process yields different outcomes depending
on the link plasticity $p$.  When $p=1$, only rewiring occurs, so that there is no effect of aging, and the results are the same as in the standard CVM: the system reaches
a frozen phase due to link rewiring. When $p=0$,  only copying occurs and it becomes the VM with aging \cite{peralta2020,gracia}.  The interevent time is determined by Eq.~\ref{eq:mean}.
This implies that, as age increases, changes in the state of the nodes become less frequent than in the original voter model. The average interevent time for copying is controlled by $\alpha_n$.
The larger the value of $\alpha_n$  the slower the dynamics of the copying process.

\section{Absorbing phase transition}

In this section, we present numerical results of the LAM and NAM models. In order to quantify the impact of aging, we measure the absolute value
of magnetization $|m|$ where the magnetization is defined as $m=(1/N) \sum_i s_i$ at a steady
state, the size of the largest connected network
component $S$, and the density of active links $\rho$ in survival runs, for various values of $\alpha$.
Here active links stand for links that connect nodes in different states, and survival runs refer to realizations of the dynamics in which, at a given time,  links remain in the network.
When all active links disappear, the system reaches an absorbing state.
The magnetization as a function of $p$ shows how the average
state of the nodes changes with copying and rewiring.
The size of the largest connected component $S$ shows,
how the structural connectivity of the network changes with different values of $p$.
Finally, the density $\rho$ of active links is a local measure of how far the system is from the absorbing state.

An active phase of the system can be identified by a finite value of $\rho$ in the thermodynamic limit.
This feature is captured in finite systems by measuring $\rho$ in survival runs.
For any finite system, a consensus phase is reached due to finite-size fluctuations, which
can be confirmed with $|m|=1$ and $S=1$.
On the other hand, the frozen phase is characterized by $\rho=0$, corresponding  to an absorbing state.
In finite systems, the frozen phase appears as a fragmented network where
the network structurally breaks into two large components with different states, resulting in $S \approx 0.5$
and $|m|=0$. A phase transition occurs between these two phases at a
critical value of plasticity, $p_c$. The mechanism driving this transition is the competition between
copying and rewiring.

\subsection{Shift of the location of absorbing transitions}

We examine the LAM and NAM focusing on the effect of aging by varying $\alpha_l$ and $\alpha_n$.
In Fig.~\ref{fig3}, the absolute value $|m|$ of magnetization,
the size $S$ of the largest connected component, and the density $\rho$ of active links
in survival runs are shown as functions of the plasticity $p$
for both the LAM in Fig.~\ref{fig3}(a-c) and the NAM in Fig.~\ref{fig3}(d-f).
We use random regular networks with the average degree $z=4$ and $N=10^4$ as initial
networks. 
In this and following sections, unless noted otherwise, we use random regular networks with the average degree $z=4$ and $N=10^4$ as initial networks.
We also verify the generality of our results starting from Erd\H{o}s-R\'enyi (ER) networks. The corresponding results are provided in the appendix and largely align with
those observed in random regular networks.
The coevolutionary dynamics starts from randomly assigned states for each node with symmetric initial states.
Active and frozen phases are observed in LAM and NAM as in the standard CVM \cite{vazquez2008}.
As discussed in the previous section, the average interevent time for the change of the state of the nodes or for link rewiring  changes
due to the effects of aging, and this leads to a shift in the critical plasticity value $p_c$.

The impact of aging appears differently in the LAM and NAM models.
In the LAM, the rewiring of links between nodes is slowed
down due to link aging. This has the effect of increasing the critical value of the plasticity, $p_c$
with increasing $\alpha_l$. A larger value of $p_c$ means that the network can maintain
its active state at higher levels of the plasticity $p$.
Thus,  aging in the LAM reduces the tendency of the dynamics to reach an absorbing
frozen state, and thereby implying larger times before reaching the consensus phase in a finite network.
On the other hand, in the NAM, aging impacts the copying dynamics.
Aging here leads to an earlier onset of the frozen phase, with smaller values of the critical plasticity $p_c$ as $\alpha_n$ increases.
In summary, aging in the CVM has opposite effects: while link aging
delays the appearance of the frozen phase and maintains the active phase by slowing link
rewiring, node aging in the NAM facilitates the existence of the frozen phase by slowing down the copying processes, thereby preventing reaching the consensus phase in finite systems.

\subsection{Absence of absorbing phase transitions}

\begin{figure}
\includegraphics[width=\linewidth]{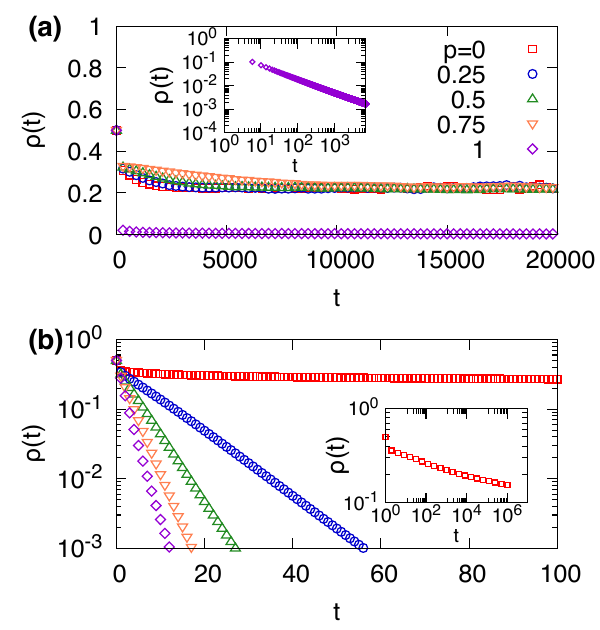}
\caption{
The density of active links $\rho(t)$ as a function of time for
(a) LAM with $\alpha_l=1$ and (b) NAM with $\alpha_n=1$ with various values
of $p$. The insets show $\rho(t)$ in a log scale for (a) LAM with $p=1$ and (b) NAM with $p=0$.
}
\label{fig4}
\end{figure}

As aging becomes stronger either by node or link aging,
the generic absorbing phase transitions between two phases may no longer occur.
The absence of absorbing phase transitions can be expected from the divergence
of the average interevent time caused by aging when $\alpha \ge 1$.
The divergence of the average interevent time implies that
link rewiring in LAM and copying in NAM are completely blocked due to aging.
Considering that the absorbing phase transition in the CVM occurs by the competition between
link rewiring and copying processes, the aging with $\alpha \ge 1$  prevents the occurrence of an
absorbing phase transition.

In the case of the LAM, as $\alpha_l$ increases, the rewiring of links between nodes slows down.
When the aging parameter $\alpha_l$ becomes one, this slowdown becomes so extreme
that the system cannot reach the frozen phase for any value of $p<1$.
This means that the system maintains the active phase because
the rewiring process is effectively suppressed by link aging.
In this scenario, an absorbing phase transition disappears, and the system
stays in an active state in a connected network across the entire parameter space with $p<1$.

We confirm the absence of the absorbing phase transition by checking the density of active links
over time, $\rho(t)$. Figure~\ref{fig4}(a) represents $\rho(t)$ for the LAM with $\alpha_l=1$ for
different values of the rewiring probability $p$. The density $\rho$ remains at a constant value
which is an evidence of the active phase for any value of $p$, except $p=1$.
When $p=1$, only link rewiring occurs and $\rho(t)$ decays with a power-law tail
as shown in the inset of Fig.~\ref{fig4}(a). Therefore, our findings confirm that the aging
in the CVM eliminates the phase transitions when $\alpha_l=1$.

In the NAM for $\alpha_n=1$, similarly to the LAM there is no
absorbing phase transition, yet in a different way.
As the aging parameter $\alpha_n$ approaches one, the copying
process slows to such an extent that $p_c$ drops to zero.
In this case, the system can no longer maintain the active phase, and leads
to the frozen phase for any value of $p>0$.
Therefore, the system does not reach a consensus in a finite system due
to the extremely slow copying process.

In Fig.~\ref{fig4}(b), corresponding to the NAM, $\rho(t)$ is plotted
for different values of $p$. In this case, it reflects the dynamics
in a network with node aging.
The decay of $\rho(t)$ follows an exponential approach to the frozen state, except in the $p=0$ case.
When $p=0$, the density decays with a power-law tail as shown in the inset of Fig.~\ref{fig4}(b), implying that the ordering dynamics becomes extremely slow. This corresponds to the ordinary Voter Model with aging \cite{gracia}.
Therefore, our findings confirm that the aging in the CVM eliminates
the phase transitions when $\alpha_n=1$.

So far we have analyzed the results for $\alpha \le 1$. For $\alpha>1$
the average interevent time diverges as in the case $\alpha =1$ due to the effect of aging,
as shown in Eq.~\ref{eq:mean}. Therefore we can expect the absence of absorbing phase transitions
regardless of $p$, to persist for $\alpha >1$.
That is, for the LAM model, the system remains in a dynamically active phase whereas
in the NAM model it reaches a frozen state, regardless of $p$.

\subsection{Phase diagram}

\begin{figure}
\includegraphics[width=\linewidth]{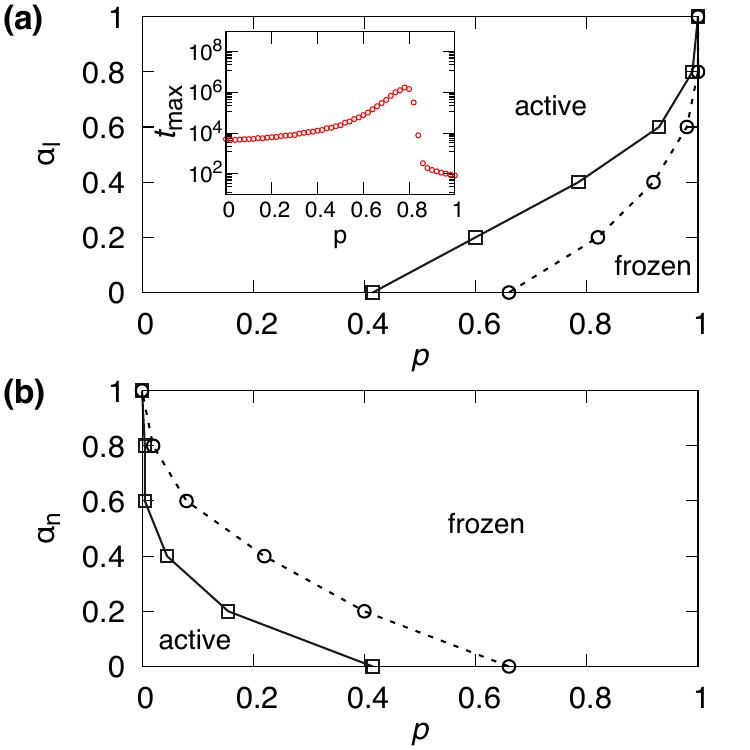}
\caption{
Phase diagram for (a) LAM and (b) NAM with active
and frozen phases as functions of network plasticity $p$
and aging parameter $\alpha$ for random regular networks
with $z=4$ (solid) and $8$ (dotted) and $N=10^4$.
The inset in Fig.~(a) shows the time $t_{max}$ to reach absorbing state for the LAM with $\alpha_l=0.4$ and $z=4$.
}
\label{fig5}
\end{figure}

The phase diagrams in Fig.~\ref{fig5} indicate the location of the phase transition between
the dynamically active and frozen phases in the parameter space of the aging parameter $\alpha$
and the plasticity $p$.
Figure \ref{fig5}(a) and (b) show respectively the phase diagram for the LAM and NAM
for random regular networks with $z=4,8$ and $N=10^4$.
The transition points are estimated in numerical simulations by the values of $p$ and $\alpha$ for which there is a maximum value of the time $t_{max}$ to reach the final state of a finite system. The inset of Fig.~\ref{fig5}(a) shows an example of the peak of $t_{max}$ at a transition point $p_c$ for $\alpha_l=0.4$ for the LAM model.

These phase diagrams highlight that the transition point between active and
frozen phases continuously varies as a function of the aging parameters
$\alpha$, showing that aging plays a critical role in the CVM.
In Fig.~\ref{fig5}(a), corresponding to the LAM, the phase transition is depicted for
various values of the link aging parameter $\alpha_l$.
As the link aging parameter $\alpha_l$ increases, the critical point $p_c$
shifts to higher values, indicating a delayed frozen state.
In Fig.~\ref{fig5}(b), for the NAM, the phase transition is similarly shown in terms of the node
aging parameter $\alpha_n$.
In the NAM model, increasing  $\alpha_n$ leads to smaller values of $p_c$, reflecting
that node aging slows down the copying process.
Note that the phase transition disappears when $\alpha_l$ or
$\alpha_n$ equals to one. When $\alpha \ge 1$, the LAM and NAM models
shows solely the active or frozen phase regardless of $p$ as
predicted by the divergence of the interevent time observed in Eq.~\ref{eq:mean}.

\subsection{Structure of rewired networks}

We studied the structure of rewired networks in the steady state by examining
the degree distribution of them. Figure~6 shows the degree distribution $P(k)$ in
the steady state for the (a) LAM and (b) NAM with $\alpha=0.4$ for various values of rewiring probability $p$, i.e., $p=0.25,0.5,1$.
The coevolutionary dynamics starts from random regular networks with average degree $z = 4$
and size $N = 10^4$. At the beginning of the coevolutionary process, the degree distribution
is a Kronecker delta function centered at $k=4$, meaning that every node in the network
has exactly four neighbors.

At the steady state, the degree distribution is changed due to link rewiring.
In LAM shown in Fig. 6(a), when $p=1$, only rewiring occurs without copying, and the dynamics end once the initially given active links are removed via rewiring. In this case, the shape of degree distribution broadens but still maintains its notable peak at $k=4$. For $p<1$, the interplay between copying and rewiring further broadens the degree distribution. 
As $p$ decreases, the system transitions from the frozen phase to the active phase, and the degree distribution $P(k)$ becomes more similar to a Poisson distribution, which is shown as a dotted line in Fig.~6. 
In the NAM as in the LAM, the degree distribution broadens as $p$ decreases. In Fig.~6(b), the frozen phase, $p=0.5$, exhibits a shape closer to a delta function, while the active phase, $p=0.1$, the distribution becomes more similar to a Poisson distribution.

\begin{figure}
\includegraphics[width=\linewidth]{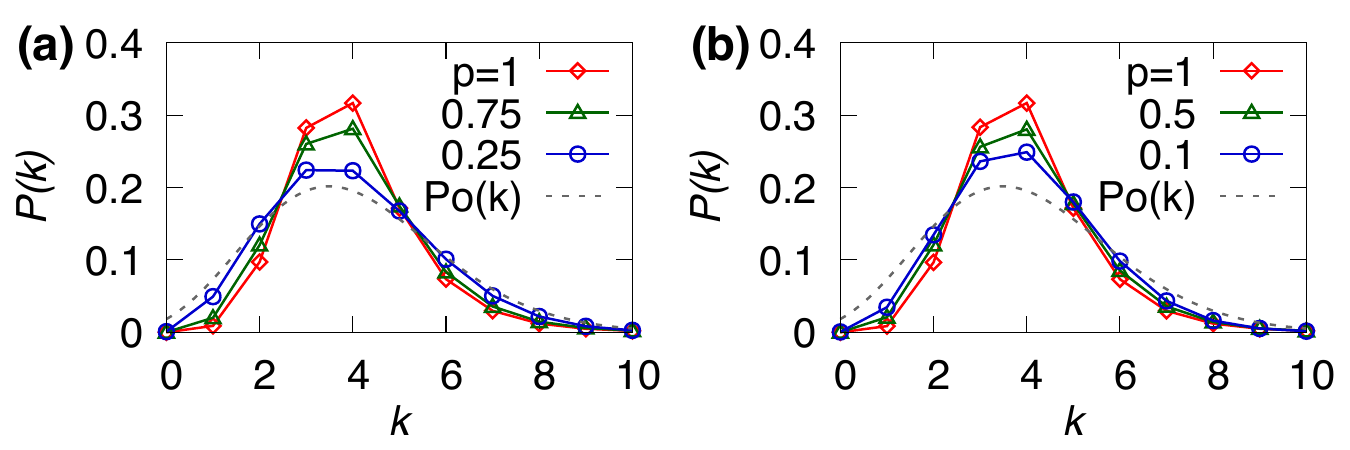}
\caption{
Rewired networks for (a) LAM with $\alpha=0.2$ and  $p=1,0.75,0.25$,
and (b) NAM with $p=1,0.5,0.1$ starting
from random regular networks with $z=4$ and $N=10^4$. The Poisson degree distribution
$Po(k)$ with $z=4$ is shown as dotted lines for comparison.
}
\label{fig6}
\end{figure}

\begin{figure*}
\includegraphics[width=\linewidth]{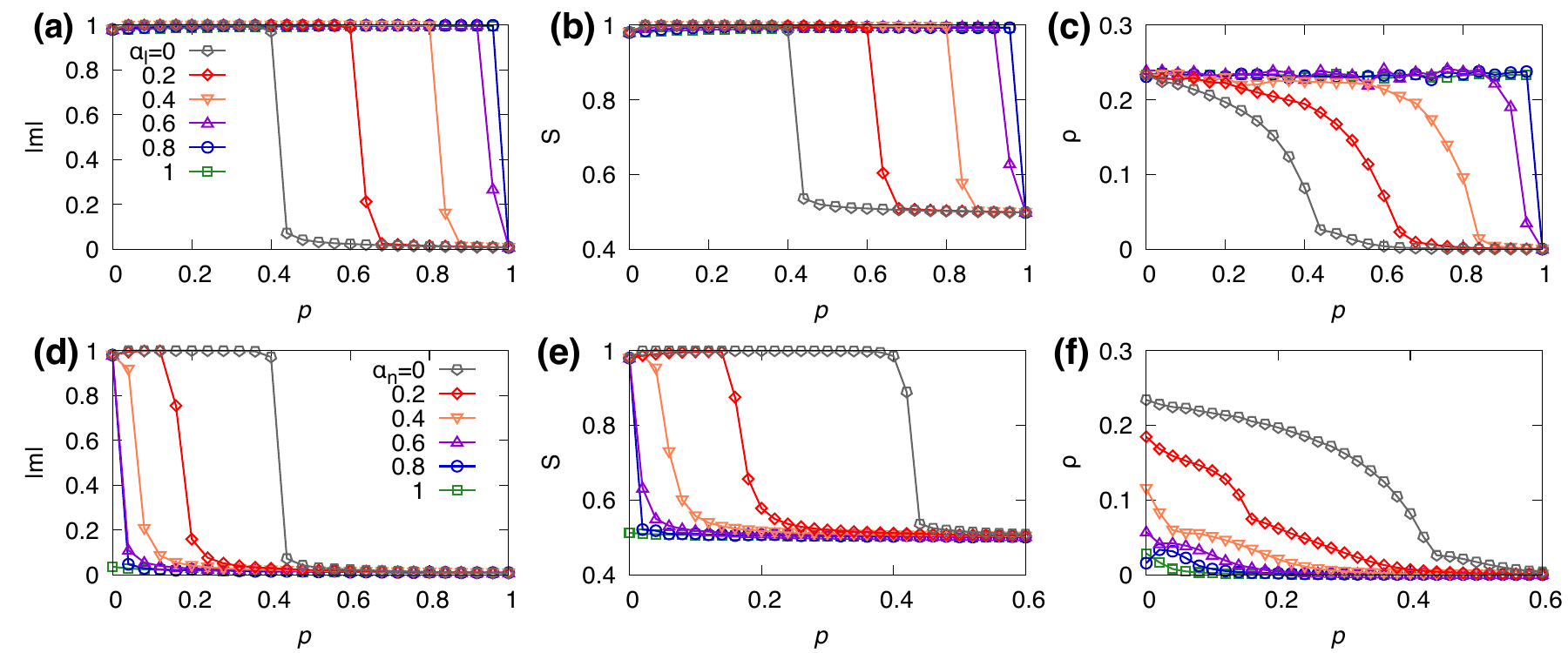}
\caption{
Results for ER networks with average degree $\langle k \rangle=4$ and $N=10^4$.
LAM:(a) The absolute value of magnetization $|m|$, (b) the size of the largest component $S$, 
(c) the fraction of active links in survival runs $\rho$ as a function of $p$ for various $\alpha_l$ for 
the link aging model, and  NAM: (d) $|m|$, (e) $S$, (f) $\rho$ for various $\alpha_n$ for 
the node aging model are shown. 
}
\label{fig7}
\end{figure*}

\section{Conclusions}

In this paper we have considered the effects of aging on the coevolving voter model (CVM), a model in which there is a coupled dynamics of the change of the states of the nodes and the time evolution of the network structure by link rewiring. We have analyzed two different forms of aging: In node aging, there is a smaller probability to change the state of the node, the longer is the persistence time of the node in that state. In link aging, the probability to rewire a link is smaller the longer that particular link has been present in the network.  We make predictions on the effect of aging in the CVM by a general analysis of the interevent time distributions. We have addressed the question of how the absorbing phase transition that occurs in the CVM for a critical value of the network plasticity $p$ is modified by aging. We find that the active and frozen phases of the standard CVM survive in the presence of aging. However, link aging increases the critical plasticity value $p_c$ for the transition to the frozen state, while node aging diminishes $p_c$, bringing
the network to the frozen phase at a lower plasticity value. For strong enough aging processes, only the active phase exists when there is link aging, while only the frozen phase survives for node aging.

Our results open the way to the study of aging effects, in particular link aging associated
with persisting links in evolving networks in a variety of different models
of coevolution studied in the literature \cite{gross2006,perc2013,raducha_spin,bmin2023}.
Understanding how aging influences the interplay between the dynamics of the structure of a complex network and the dynamics of the states of the nodes of the network
could provide insights in the evolution of many complex systems.
However, an open question remains for future work about the competition and possible
synergistic effects of different forms of aging, such as aging of nodes and links.

\section{Acknowledgments}

This work was supported by National Research Foundation of Korea (NRF) grants funded by the Korea government (MSIT) (No. 2020R1I1A3068803) and
Global Learning \& Academic research institution for Master’s and PhD students, and Postdocs(LAMP) Program of the National Research Foundation of Korea (NRF) grant funded by the Ministry of Education (No. RS-2024-00445180) (BM), and by the Agencia Estatal de Investigaci\'on (AEI, MICIU, Spain) MICIU/AEI/10.13039/501100011033 and Fondo Europeo de Desarrollo Regional (FEDER, UE) under Project APASOS (PID2021-122256NB-C21) and the Mar\'ia de Maeztu Program for units of Excellence in R\&D, grant CEX2021-001164-M (MSM).

\appendix

\section{ER networks}

We examine the LAM and NAM starting from ER networks to confirm the generality of our findings.
Similar to random regular networks, we study the effects of aging by varying $\alpha_l$
and $\alpha_n$. In both models, we measure the absolute value $|m|$ of magnetization,
the size $S$ of the largest connected component, and the density $\rho$ of active links in survival runs
as functions of the link rewiring parameter $p$. 
The results show the transition between active and frozen phases, as observed in the standard CVM.

The effects of aging in ER networks shown in Fig.~7 follow the same qualitative patterns as 
in random regular networks. 
That is, the aging mechanism shifts the critical plasticity $p_c$.
In the LAM, link aging slows down rewiring, increasing $p_c$ and extending the active phase.
Conversely, in the NAM, node aging slows the copying process, lowering $p_c$
and leading to an earlier transition to the frozen phase. The results for ER networks in this appendix
align qualitatively with those observed in random regular networks.

\end{document}